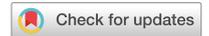

# Fluctuation relations and fitness landscapes of growing cell populations

Arthur Genthon✉ & David Lacoste

We construct a pathwise formulation of a growing population of cells, based on two different samplings of lineages within the population, namely the forward and backward samplings. We show that a general symmetry relation, called fluctuation relation relates these two samplings, independently of the model used to generate divisions and growth in the cell population. These relations lead to estimators of the population growth rate, which can be very efficient as we demonstrate by an analysis of a set of mother machine data. These fluctuation relations lead to general and important inequalities between the mean number of divisions and the doubling time of the population. We also study the fitness landscape, a concept based on the two samplings mentioned above, which quantifies the correlations between a phenotypic trait of interest and the number of divisions. We obtain explicit results when the trait is the age or the size, for age and size-controlled models.

While the growth of cell populations appears deterministic, many processes occurring at the single cell level are stochastic. Among many possibilities, stochasticity at the single cell level can arise from stochasticity in the generation times[1], from stochasticity in the partition at division[2,3], or from the stochasticity of single cell growth rates, which are usually linked to stochastic gene expression[4]. Ideally one would like to be able to disentangle the various sources of stochasticity present in experimental data[5]. This would allow to understand and predict how the various sources of stochasticity affect macroscopic parameters of the cell population, such as the Malthusian population growth rate[6,7]. Beyond this specific question, research in this field attempts to elucidate the fundamental physical constraints which control growth and divisions in cell populations.

With the advances in single cell experiments, where the growth and divisions of thousand of individual cells can be tracked, robust statistics can be acquired. New theoretical methods are needed to exploit this kind of data and to relate experiments carried out at the population level with experiments carried out at the single cell level. For instance, one would like to relate single-cell time-lapse video microscopy experiments of growing cell populations[8], which provide information on all the lineages in the branched tree, with experiments carried out with the mother machine configuration, which provide information on single lineages[9,10].

Let us now review quickly how the issue was addressed theoretically. In 2015, a pathwise thermodynamic framework was built for population dynamics using large deviation theory. One important result was a variational principle for the population growth rate[11], which was formulated in terms of two key path distributions, namely the chronological and the retrospective probability distributions. Then, in order to explain their experimental observation that populations of *Escherichia coli* double faster than the mean doubling time of their constituent single cells, Hashimoto et al. extended the classical work of Powell[6] for age models without mother-daugher correlations[12]. Nozoe et al.[13] then showed that the difference between the forward (chronological) and backward (retrospective) distributions can be used to define a quantity called phenotypic fitness landscape, which informs whether a specific phenotypic trait affects the population growth rate. In that work, they also derived the key relation between the two distributions, already known in the mathematical literature of branching processes[14], but they did not connect this result with the field of fluctuation relations. Various theoretical works followed which addressed other aspects of the role of the stochasticity at the single cell level[3,15,16]. As far as we can tell, the connection between the results of Nozoe and the field of fluctuation relations was only made explicit in our first work on this topic[17]. In that work, we also derived inequalities for mean generation times, already obtained in[12] for age models, but importantly we proved using the fluctuation relation that they are valid beyond age models, in particular for a broad class of size models.

Gulliver, CNRS, ESPCI Paris, PSL University, 75005 Paris, France. ✉email: arthur.genthon@hotmail.fr

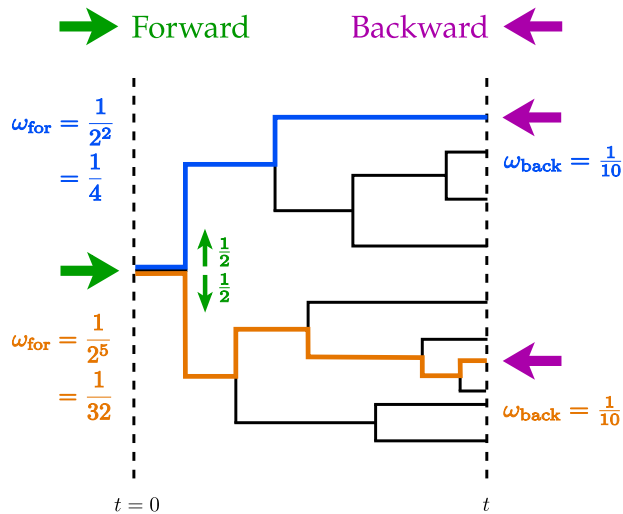

**Figure 1.** Example of a tree with $N_0 = 1$ and $N(t) = 10$ lineages at time $t$. Two lineages are highlighted, the first in blue with 2 divisions and the second in orange with 5 divisions. The forward sampling is represented with the green right arrows: it starts at time $t = 0$ and goes forward in time by choosing one of the two daughters lineages at each division with probability 1/2. The backward sampling is pictured by the left purple arrows: starting from time $t$ with uniform weight on the 10 lineages it goes backward in time down to time $t = 0$.

In the present work, we further investigate the connection between the statistics at the single lineage and population levels using fluctuation relations. These fluctuation relations only depend on the structure of the branched tree but not on the class of dynamical variables (protein concentration, cell size or cell age.) defined on it. These relations imply that the above inequalities for the mean number of divisions or mean generation times hold regardless of the specific dynamics.

We then provide an interpretation of the fluctuation relations within Stochastic Thermodynamics[18] and we explore some consequences. One application concerns the inference of the population growth rate using single lineage data[19]. We then introduce some specific dynamical models, which we simulate to confirm the fluctuation relations and their consequences for mean generation times. Then, for these specific models and for key phenotypic variables such as the size and the age, we also study the fitness landscape[13].

## Theoretical framework

**The backward and forward processes.** Let us consider a branched tree, starting with $N_0$ cells at time $t = 0$ and ending with $N(t)$ cells at time $t$ as shown on Fig. 1. We assume that all lineages survive up to time $t$, and therefore the final number $N(t)$ of cells corresponds to the number of lineages in the tree.

The most natural way to sample the lineages is to put uniform weights on all of them. This sampling is called backward, (or retrospective) because at the end of the experiment one randomly chooses one lineage among the $N(t)$ with a uniform probability and then one traces the history of the lineage backward in time from time $t$ to 0, until reaching the ancestor population. The backward weight associated with a lineage $l$ is defined as

$$\omega_{\text{back}}(l) = N(t)^{-1}. \quad (1)$$

In a tree, some lineages divide more often than others, which results in an over-representation of lineages that have divided more often than the average. Therefore by choosing a lineage with uniform distribution, we are more likely to choose a lineage with more divisions than the average number of divisions in the tree.

The other way of sampling a tree is the forward (or chronological) one and consists in putting the weight

$$\omega_{\text{for}}(l) = N_0^{-1} m^{-K(l)}, \quad (2)$$

on a lineage $l$ with $K(l)$ divisions, where $m$ is the number of offspring at division. This choice of weights is called forward because one starts at time 0 by uniformly choosing one cell among the $N_0$ initial cells, and one goes forward in time up to time $t$, by choosing one of the $m$ offspring with equal weight $1/m$ at each division. The backward and forward weights are properly normalized probabilities, defined on the $N(t)$ lineages in the tree at time $t$: $\sum_{i=1}^{N(t)} \omega_{\text{back}}(l_i) = \sum_{i=1}^{N(t)} \omega_{\text{for}}(l_i) = 1$.

Single lineage experiments are precisely described by a forward process since experimentally, at each division, only one of the two daughter cells is conserved while the other is eliminated (for instance flushed away in a microfluidic channel[9, 10]). In these experiments, a tree is generated but at each division only one of the two lineages is conserved, with probability 1/2, while the rest of the tree is eliminated. This means that single lineage observables can be measured without single lineage experiments, provided population experiments are analyzed with the correct weights on lineages.

**Link with the population growth rate.** Since the backward weight put on a lineage depends on the number of cells at time *t*, it takes into account the reproductive performance of the colony but it is unaffected by the reproductive performance of the lineage considered. On the contrary, the forward weight put on a specific lineage depends on the number of divisions of that lineage but is unaffected by the reproductive performance of other lineages in the tree. Therefore, the difference between the values of the two weights for a particular lineage informs on the difference between the reproductive performance of the lineage with respect to the colony.

We now introduce the population growth rate:

$$\Lambda_t = \frac{1}{t} \ln \frac{N(t)}{N_0}, \tag{3}$$

which is linked to forward weights by the relation

$$\frac{N(t)}{N_0} = \sum_{i=1}^{N(t)} m^{K_i} \omega_{\text{for}}(l_i) = \langle m^K \rangle_{\text{for}}, \tag{4}$$

where $\langle \cdot \rangle_{\text{for}}$ is the average over the lineages weighted by $\omega_{\text{for}}$, and $K_i = K(l_i)$. Combining the two equations above, we obtain[19]:

$$\Lambda_t = \frac{1}{t} \ln \langle m^K \rangle_{\text{for}}, \tag{5}$$

which allows an experimental estimation of the population growth rate from the knowledge of the forward statistics only.

Equation (4) can also be re-written to express the bias between the forward and backward weights of the same lineage

$$\frac{\omega_{\text{back}}(l)}{\omega_{\text{for}}(l)} = \frac{m^{K(l)}}{\langle m^K \rangle_{\text{for}}}, \tag{6}$$

which is the reproductive performance of the lineage divided by its average in the colony with respect to $\omega_{\text{for}}$.

A similar relation is derived using the relation

$$\frac{N_0}{N(t)} = \sum_{i=1}^{N(t)} m^{-K_i} \omega_{\text{back}}(l_i) = \langle m^{-K} \rangle_{\text{back}}. \tag{7}$$

Combining Eqs. (5) and (7) we obtain:

$$\Lambda_t = -\frac{1}{t} \ln \langle m^{-K} \rangle_{\text{back}}. \tag{8}$$

A similar equation as Eq. (6) can be obtained in terms of the backward sampling and reads:

$$\frac{\omega_{\text{back}}(l)}{\omega_{\text{for}}(l)} = \frac{\langle m^{-K} \rangle_{\text{back}}}{m^{-K(l)}}. \tag{9}$$

Combining Eqs. (1) to (3), we obtain the fluctuation relation[13,17]:

$$\omega_{\text{back}}(l) = \omega_{\text{for}}(l) \, e^{K(l) \ln m - t \Lambda_t}. \tag{10}$$

If we now introduce the probability distribution of the number of divisions for the forward sampling $p_{\text{for}}(K) = \sum_l \delta(K - K(l)) \omega_{\text{for}}(l)$ and similarly for the backward sampling, we can also recast the above relation as a fluctuation relation for the distribution of the number of divisions:

$$p_{\text{back}}(K, t) = p_{\text{for}}(K, t) \, e^{K \ln m - t \Lambda_t}. \tag{11}$$

Let us now introduce the Kullback–Leibler divergence between two probability distributions *p* and *q*, which is the non-negative number:

$$\mathscr{D}_{\text{KL}}(p||q) = \int dx \, p(x) \ln \frac{p(x)}{q(x)} \geq 0. \tag{12}$$

Using Eq. (10), we obtain

$$\mathscr{D}_{\text{KL}}(\omega_{\text{back}}||\omega_{\text{for}}) = \langle K \rangle_{\text{back}} \ln m - t \Lambda_t \geq 0. \tag{13}$$

A similar inequality follows by considering $\mathscr{D}_{\text{KL}}(\omega_{\text{for}}||\omega_{\text{back}})$. Finally we obtain

$$\frac{t}{\langle K \rangle_{\text{back}}} \leq \frac{\ln m}{\Lambda_t} \leq \frac{t}{\langle K \rangle_{\text{for}}}. \tag{14}$$

In the long time limit, $\lim_{t \to +\infty} t / \langle K \rangle_{\text{back}} = \langle \tau \rangle_{\text{back}}$, where $\tau$ is the inter-division time, or generation time, defined as the time between two consecutive divisions on a lineage. The same argument goes for the forward average. In the case of cell division where each cell only gives birth to two daughter cells ($m = 2$), the center term in the inequality tends to the population doubling time $T_d$. Therefore, this inequality reads in the long time limit:

$$\langle \tau \rangle_{\text{back}} \leq T_d \leq \langle \tau \rangle_{\text{for}}. \tag{15}$$

Let us now mention a minor but subtle point related to this long time limit. For a lineage with $K$ divisions up to time $t$, we can write $t = a + \sum_{i=1}^{K} \tau_i$, where $a$ is the age of the cell at time $t$ and where $\tau_i$ is the generation time associated with the $i$th division. Then $t/K = \tau_m + a/K$, where $\tau_m$ is the mean generation time along the lineage. For finite times, all we can deduce is $t/K \geq \tau_m$. Therefore the left inequality of Eq. (15) always holds

$$\langle \tau \rangle_{\text{back}} \leq \frac{t}{\langle K \rangle_{\text{back}}} \leq \frac{\ln m}{\Lambda_t}, \tag{16}$$

while the right inequality does not necessarily hold at finite time.

Inspired by work by Powell[6], the inequalities of Eq. (15) have been theoretically derived in[12] for age models. In our previous work[17], we have replotted the experimental data of[12] which confirm theses inequalities and we have shown theoretically that the same inequalities should also hold for size models. In fact, as the present derivation shows, the relation equation (14) is very general and only depends on the branching structure of the tree, while the relation equation (15) requires in addition the existence of a steady state. These inequalities and Eq. (11) express fundamental constraints between division and growth, which should hold for any model.

**Stochastic thermodynamic interpretation.** The results derived above have a form similar to that found in Stochastic Thermodynamics[18]. According to this framework, Eq. (5) is an integral fluctuation relation (similar to Jarzynski relation) while Eq. (11) is a detailed fluctuation relation (similar to Crooks fluctuation relation). Furthermore, the inequalities equation (14) represent a constraint equivalent to the second law of thermodynamics, which classically follows from the Jarzynski or Crooks fluctuation relations. It is known that these inequalities take a slightly different form when expressed at finite time or at steady state, which is indeed the case here when comparing Eq. (14) with Eq. (15). A difference between work fluctuation relations like Crooks or Jarzynski and equations (5) and (11), is that Crooks or Jarzynski describe non-autonomous systems which are driven out of equilibrium by the application of a time-dependent protocol, whereas the relations for cell growth derived here concern autonomous systems, in the absence of any external protocol.

One of the main applications of Jarzynski or Crooks fluctuation relations concerns the thermodynamic inference of free energies from non-equilibrium fluctuations. Similarly, Eq. (5) or Eq. (11) can be used as estimators of the population growth rate. The specific advantage of Eq. (5) with respect to Eq. (11) is that it only requires single lineage statistics, which can be obtained from mother machine experiments. Let us now show how this can be done in practice. We use the data from[20], where the growth of many independent lineages of *E. coli* have been recorded over 70 generations in a mother machine at three different temperatures (25 °C, 27 °C, and 37 °C), precisely 65 lineages for 25 °C, 54 for 27 °C, and 160 for 37 °C. For each temperature condition, we study the convergence of the estimator of the population growth rate based on Eq. (5), which we call $\Lambda_{\text{lin}}$ as a function of the length $t$ of the lineages for a fixed number of independent lineages $L$, and as a function of the number of independent lineages for a fixed observation time.

Firstly, for each temperature, we take into account all the lineages available and truncate them at an arbitrary time $t$ smaller than the length of the shortest lineage of the set. On these portions of lineages of length $t$, we compute $\Lambda_{\text{lin}}$ versus the time $t$ as shown in Fig. 2a. We see that the estimator $\Lambda_{\text{lin}}$ starts from zero, increases and eventually converges rather quickly towards a limiting value. The limit we found agree with the independent analysis carried out in[19], with only one caveat, these authors reported that their estimator started at high values and then decreased towards the limit, while in our case, the estimator starts at zero and later increases towards the limit. In our case, the estimator needs to be zero at short times, before the first divisions occur.

Secondly, we truncate all the lineages at a fixed time equal to the length of the shortest lineage of the set, and compute $\Lambda_{\text{lin}}$ versus the number $L$ of lineages considered for the estimation, which have been randomly selected from the ensemble of available lineages. As shown in Fig. 2b for the case at 37 °C (curves for the other temperatures look exactly the same), the convergence is also excellent in that case. Although the value of the population growth rate which is obtained in this way can not be measured independently from the evolution of the population in the mother machine setup, this convergence is indicative of the success of the method. The figure also confirms that the value of the population growth rate deduced from the estimator $\Lambda_{\text{lin}}$ is larger than $\ln(2)/\langle \tau \rangle_{\text{for}}$, as predicted by the right inequality of Eq. (15).

Here, the estimator is found to provide an excellent estimation, but this is not always so. For instance, for the inference of free energies from non-equilibrium work measurements, the exponential average of the estimator is often dominated by rare values, which are not accessible or not well sampled[21]. To understand why this problem does not arise here, we show in inset of Fig. 2b, the distribution $P(K)$ of the number of divisions together with the same distribution weighted by the factor $2^K$ and normalized. The peak of that modified distribution informs on the dominant values in the estimator[21]. Here, we observe that both distributions have a narrow support and are close to each other. The weighted distribution is peaked at $K = 67$ while $P(K)$ is peaked at $K = 66$, therefore typical and dominating values are very close, which explains why the estimator is good.

Let us now further develop the Stochastic Thermodynamic interpretation of our results by analyzing the implications of the previous fluctuation relations when dynamical variables are introduced on the branched tree of the population. Let us introduce $M$ variables labeled ($y_1, y_2, \ldots, y_M$) to describe a dynamical state of the system, then a path is fully determined by the values of these variables at division, and the times of each division. We call $\mathbf{y}(t) = (y_1(t), y_2(t), \ldots, y_M(t))$ a vector state at time $t$ and $\{\mathbf{y}\} = \{\mathbf{y}(t_j)\}_{j=1}^K$ a path with $K$ divisions. For cell growth models, the variables $y_i$ can typically be the size and age of the cell, or the concentration of a key protein.

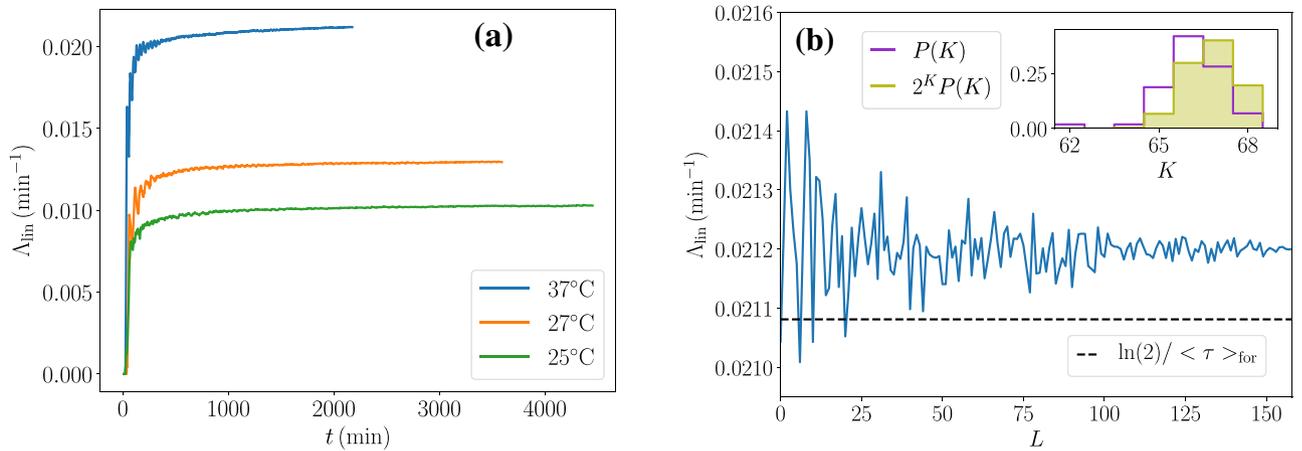

**Figure 2.** Estimator of the population growth rate $\Lambda_{\text{lin}}$ based on Eq. (5), (**a**) as function of the the length $t$ of the lineages and (**b**) as function of the number $L$ of lineages used in the estimation. In (**a**), the curves for the three temperatures converge to a constant value. In (**b**), only the curve for 37 °C is shown and the horizontal dashed line represents the quantity $\ln(2)/\langle\tau\rangle_{\text{for}}$, which is smaller than the limit value of $\Lambda_{\text{lin}}$, as expected from the second law-like inequality, namely Eq. (15). In the inset, the purple histogram is the distribution of the number of divisions, while the green filled histogram is the histogram deduced from it by weighting it by a factor $2^K$ and normalizing. All the 160 lineages were used to plot these histograms.

The probability $\mathscr{P}$ of path $\{\mathbf{y}\}$ is defined as the sum over all lineages of the weights of the lineages that follow the path $\{\mathbf{y}\}$:

$$\mathscr{P}(\{\mathbf{y}\}, K, t) = \sum_{i=1}^{N(t)} \omega(l_i)\, \delta(K - K_i)\delta(\{\mathbf{y}\} - \{\mathbf{y}\}_i), \qquad (17)$$

where $\{\mathbf{y}\}_i$ is the path followed by lineage $l_i$. Using the normalization of the weights $\omega$ on the lineages, we show that $\mathscr{P}$ is properly normalized: $\int d\{\mathbf{y}\} \sum_K \mathscr{P}(\{\mathbf{y}\}, K, t) = 1$. We then define the number $n(\{\mathbf{y}\}, K, t)$ of lineages in the tree at time $t$ that follow the path $\{\mathbf{y}\}$ with $K$ divisions:

$$n(\{\mathbf{y}\}, K, t) = \sum_{i=1}^{N(t)} \delta(K - K_i)\delta(\{\mathbf{y}\} - \{\mathbf{y}\}_i). \qquad (18)$$

This number of lineages is normalized as $\int d\{\mathbf{y}\} \sum_K n(\{\mathbf{y}\}, K, t) = N(t)$. Then, the path probability can be rewritten as

$$\mathscr{P}(\{\mathbf{y}\}, K, t) = n(\{\mathbf{y}\}, K, t) \cdot \omega(l). \qquad (19)$$

Since $n(\{\mathbf{y}\}, K, t)$ is independent of a particular choice of lineage weighting, we obtain

$$\frac{\mathscr{P}_{\text{back}}(\{\mathbf{y}\}, K, t)}{\mathscr{P}_{\text{for}}(\{\mathbf{y}\}, K, t)} = \frac{\omega_{\text{back}}(l)}{\omega_{\text{for}}(l)} = e^{K \ln m - t\Lambda_t}, \qquad (20)$$

which generalizes Eq. (11). In our previous work[17], we have derived this relation for size models with individual growth rate fluctuations (i.e. $\mathbf{y} = (x, \nu)$) but we were not aware of the weighting method introduced by[13], and for this reason, we used the term 'tree' to denote the backward sampling, and the term 'lineage' to denote the forward sampling.

This relation has a familiar form in Stochastic Thermodynamics. The central quantity called entropy production can indeed be expressed similarly as the relative entropy between probability distributions associated with a forward and a backward evolution. In this analogy, $\{\mathbf{y}\}$ is analog to the trajectory and $t\Lambda_t - K \ln m$ is analog to the entropy production. Then, the equivalent of a reversible trajectory for which the entropy production is null is a lineage for which the number $K$ of divisions is equal to $t\Lambda_t/\ln m$, that is, a lineage having the same reproductive performance as that of the colony. When all the lineages in a tree have this property, there is no variability of the number of divisions among them. In that case, the forward and backward distributions are identical, and the cost function $t\Lambda_t - K \ln m$ vanishes for all lineages.

### Mixed age-size controlled models

**Dynamics at the population level.** The state of a cell is described by its size $x$, its age $a$ and its individual growth rate $\nu$, with $\mathbf{y} = (x, a, \nu)$. Such mixed size-age model includes the 'adder' in which the cell divides after adding a constant volume to its birth volume[22–25].

The evolution of the number of cells $n(\mathbf{y}, K, t)$ in the state $\mathbf{y}$ at time $t$, that belong to a lineage with $K$ divisions up to time $t$ is governed by the equation

$$(\partial_t + \partial_a) n(\mathbf{y}, K, t) + \partial_x [\nu x n(\mathbf{y}, K, t)] + B(\mathbf{y}) n(\mathbf{y}, K, t) = 0, \tag{21}$$

and the boundary condition

$$n(x, a = 0, \nu, K, t) = m \int d\mathbf{y}' B(\mathbf{y}') \Sigma(\mathbf{y}|\mathbf{y}') n(\mathbf{y}', K-1, t), \tag{22}$$

where $B(\mathbf{y})$ is the division rate and $\Sigma(\mathbf{y}|\mathbf{y}')$ is the conditional probability (also called division kernel) for a newborn cell to be in state $\mathbf{y}$ knowing its mother divided while in state $\mathbf{y}'$, normalized as $\int \Sigma(\mathbf{y}|\mathbf{y}') d\mathbf{y} = 1$, for any $\mathbf{y}'$.

**Dynamics at the probability level.** While $n(\mathbf{y}, K, t)$ in Eq. (21) is independent of the choice of weights put on the lineages, we now turn to a description in terms of the probability $p(\mathbf{y}, K, t)$ for a cell to be in state $(\mathbf{y}, K)$ at time $t$ if chosen randomly among the $N(t)$ cells in the tree at that time. To do so, one has to choose how to weight each cell in the colony, which is equivalent to weight each lineage, since at time $t$ each cell is the ending point of one lineage.

The first possibility is the backward sampling, for which each lineage is weighted uniformly. In this case, we define $p_{\text{back}}$ as

$$p_{\text{back}}(\mathbf{y}, K, t) = \frac{n(\mathbf{y}, K, t)}{N(t)}. \tag{23}$$

Dividing Eq. (21) and the boundary condition equation (22) by $N(t)$ we obtain

$$(\partial_t + \partial_a) p_{\text{back}}(\mathbf{y}, K, t) + \partial_x [\nu x p_{\text{back}}(\mathbf{y}, K, t)] + [B(\mathbf{y}) + \Lambda_p(t)] p_{\text{back}}(\mathbf{y}, K, t) = 0, \tag{24}$$

and

$$p_{\text{back}}(x, a = 0, \nu, K, t) = m \int d\mathbf{y}' B(\mathbf{y}') \Sigma(\mathbf{y}|\mathbf{y}') p_{\text{back}}(\mathbf{y}', K-1, t), \tag{25}$$

where we defined the instantaneous population growth rate as

$$\Lambda_p(t) = \frac{\dot{N}}{N}. \tag{26}$$

The instantaneous population growth rate and the population growth rate defined in Eq. (3) are related by:

$$\Lambda_t = \frac{1}{t} \int_0^t \Lambda_p(t') dt'. \tag{27}$$

In the long-time limit, $N$ grows exponentially with constant rate $\Lambda_p$, and thus $\Lambda_t = \Lambda_p = \Lambda$.

The other possibility is to use the forward statistics, in which case we define the probability $p_{\text{for}}$ as

$$p_{\text{for}}(\mathbf{y}, K, t) = \frac{n(\mathbf{y}, K, t)}{m^K}. \tag{28}$$

Dividing Eq. (21) and the boundary condition equation (22) by $m^K$ we obtain

$$(\partial_t + \partial_a) p_{\text{for}}(\mathbf{y}, K, t) + \partial_x [\nu x p_{\text{for}}(\mathbf{y}, K, t)] + B(\mathbf{y}) p_{\text{for}}(\mathbf{y}, K, t) = 0, \tag{29}$$

and

$$p_{\text{for}}(x, a = 0, \nu, K, t) = \int d\mathbf{y}' B(\mathbf{y}') \Sigma(\mathbf{y}|\mathbf{y}') p_{\text{for}}(\mathbf{y}', K-1, t). \tag{30}$$

One can notice that the backward statistics is well suited to study the population, while the forward statistics reproduce the behaviour of single lineage experiments. Indeed, by taking Eqs. (24) and (25) for the population/backward probability $p_{\text{back}}$, and choosing $\Lambda_p(t) = 0$ and $m = 1$ we recover Eqs. (29) and (30). This equation is then a population equation in which we follow only one cell, so that $\Lambda_p(t) = 0$ and $m = 1$, which we call single lineage experiment.

**Illustration of the fluctuation relation.** We simulated the time evolution of colonies of cells, obeying Eqs. (21) and (22), for age and size models in order to illustrate the fluctuation relation. Since results are very similar—as expected—for age models, we restrict ourselves to size models. We tested two results: the fluctuation relation for the number of divisions Eq. (11) and one of its consequences: the inequality for the mean number of divisions Eq. (14).

All simulations for size models were conducted with the division rate $B(x, \nu) = \nu x^\alpha$, where $\alpha$ is the strength of the control and $x$ is the dimensionless size. Power law were found to be good approximations for empirical division rates $B(x)$[2,24,26]. The factor $\nu$, being the only time scale for size models, gives $B(x)$ its proper dimension. Similarly for age models[26], we choose $B(a, \nu) = \nu a^\alpha$.

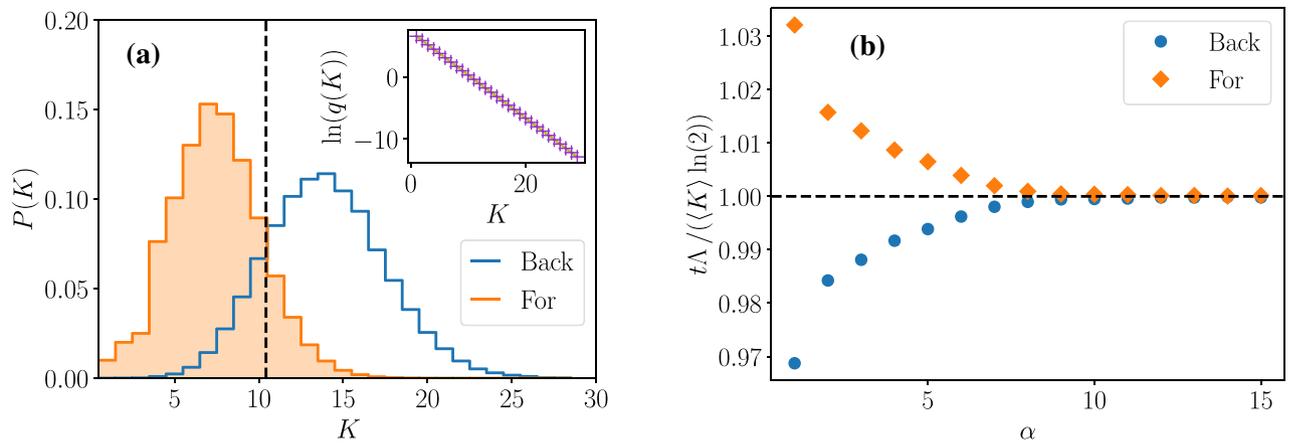

**Figure 3.** Illustration of (**a**) the fluctuation relation for the number of divisions and (**b**) the inequality on the mean number of divisions, for a size-controlled model with division rate $B(x) = \nu x^\alpha$. In (**a**), the forward distribution is shown as orange filled histogram and the backward distribution as blue empty histogram. The vertical dashed line at $K = t\Lambda_t / \ln 2$ is the theoretical value at which the two distributions should intersect. The inset shows the logarithm of the ratio $q(K)$ of the forward to backward probabilities (purple crosses), and the theoretical result $t\Lambda_t - K \ln 2$ (green line). The conditional probability $\Sigma(x|x')$ is uniform between sizes 0 and $x'$. The simulation was conducted with constant $\nu = 1$, $\alpha = 2$ and $t = 7$. In (**b**), the quantity $t/\langle K \rangle$, re-scaled by the population doubling time $T_d = \ln 2 / \Lambda_t$, is plotted when $\alpha$ is varied from 1 to 15, with orange diamonds for the forward sampling and with blue circles for the backward sampling. The volume partition between the two daughter cells at division is symmetrical, so that $\Sigma(x|x') = \delta(x - x'/2)$. The simulation was conducted with constant $\nu = 1$, and $t = 6$.

On Fig. 3a, the backward and forward probability distributions of the number of divisions are shown for a size model. The two distributions intersect at the number of divisions $K = t\Lambda_t / \ln 2$. The inset of Fig. 3a shows the logarithm of the ratio $q(K, t) = p_{\text{for}}(K, t)/p_{\text{back}}(K, t)$ of the two distributions, which is as expected a straight line of slope $-\ln 2$ when plotted against the number of divisions. For convenience and for Fig. 3a only, noise in the volume partition at division has been introduced, by choosing for the conditional probability $\Sigma(x|x')$ a uniform distribution between sizes $x = 0$ and $x = x'$. This has the effect of broadening the distributions $P(K)$ with respect to the case of deterministic symmetrical volume partition.

Then, we tested the inequality on the mean numbers of divisions by varying the strength of the size-control $\alpha$. Results are shown on Fig. 3b. One one hand, we see that the less control on size, the more discrepancy between the two determinations $\langle K \rangle_{\text{back}}$ and $\langle K \rangle_{\text{for}}$. On the other hand, when increasing the control, the two determinations converge to the population doubling time, where no stochasticity in the number of divisions is left, and every lineage carries the same number of divisions, leading to the equality of the backward and forward statistics.

### Phenotypic fitness landscapes

The fitness of a phenotypic trait $s$ is a measure of the reproductive success of individuals carrying it. It is usually defined as the number of offsprings of one individual with a given value of the trait and is quite difficult to evaluate. Nozoe et al. suggested that one way to measure it could be to compare the chronological and retrospective marginal probabilities[13] and accordingly defined it as:

$$h(s) = \Lambda_t + \frac{1}{t} \ln \left[ \frac{P_{\text{back}}(s)}{P_{\text{for}}(s)} \right], \quad (31)$$

so that

$$P_{\text{back}}(s) = P_{\text{for}}(s) \exp\left[(h(s) - \Lambda_t)t\right]. \quad (32)$$

This has again the form of a fluctuation relation similar to Eq. (11), except for the replacement of the factor $K \ln 2/t$ by the function $h(s)$. This suggests that the fitness landscape $h(s)$ plays a role similar to that of an effective division rate, which depends on the trait $s$. In line with this interpretation, in the particular case where $s = K$, Eq. (11) leads to $\tilde{h}(K) = K \ln 2/t$, where the fitness landscape for trait $K$ is called the lineage fitness and is written $\tilde{h}$. In a branched tree, lineages with a large number of divisions $K$ are exponentially over-represented in the population with the backward sampling as compared to the forward sampling. This means that lineages with large $K$ have a larger fitness than the ones with a small $K$, which is coherent with $\tilde{h}(K)$ being an increasing function of $K$.

In the following, we rewrite the definition of $h(s)$ in a slightly different way[17] using

$$P_{\text{back}}(s) = e^{-t\Lambda_t} P_{\text{for}}(s) \sum_K 2^K R_{\text{for}}(K|s), \quad (33)$$

where we have introduced the probability of the number of division events conditioned on trait $s$ at the forward level, $R_{\text{for}}(K|s)$. Lastly, the fitness landscape reads[17]

$$h(s) = \frac{1}{t} \ln \left[ \sum_K 2^K R_{\text{for}}(K|s) \right]. \tag{34}$$

An increasing or decreasing fitness landscape means a positive or negative correlation of the trait value with the capacity to divide, whereas a constant fitness landscape means that the trait is not correlated with the number of divisions. Indeed, if we consider a trait $s$ which does not affect the number $K$ of divisions, then $R_{\text{for}}(K|s) = P_{\text{for}}(K)$ and Eq. (34) reads $h(s) = \ln \left[ \sum_K 2^K P_{\text{for}}(K) \right]/t$, which is equal to $\Lambda_t$ according to Eq. (5). In that case, we find that the backward and forward probabilities for that trait $s$ are equal.

In the next sections, we evaluate the relevance of the key variables from our model, namely the size and the age by evaluating their fitness landscapes in size and age models.

**Size models.** We start with a case where the fitness landscape is fully solvable namely a size model with no individual growth rate fluctuations and with symmetric division. Let us consider a colony starting with one ancestor cell of size $x_0$. Then, the available sizes at time $t$ are discrete and given by $x = x_0 \exp[\nu t]/2^K$ where $K$ is the number of divisions undergone by the cell. Therefore a particular size $x$ can be reached only if there is an integer $K$ satisfying this relation, and this integer is unique, leading to

$$R_{\text{for}}(K|x) = \delta \left( K - \frac{\ln \left[ \frac{x_0 e^{\nu t}}{x} \right]}{\ln 2} \right). \tag{35}$$

Using this relation in Eq. (34), one finds

$$h(x) = \nu + \frac{1}{t} \ln \left( \frac{x_0}{x} \right). \tag{36}$$

The fitness landscape of the size is a decreasing function, which is coherent with the over-representation of cells that divided a lot, since these cells are more likely to be small due to the numerous divisions. Reporting this result in Eq. (33), we obtain a fluctuation relation for the size

$$P_{\text{back}}(x) = e^{(\nu - \Lambda_t)t} \frac{x_0}{x} P_{\text{for}}(x), \tag{37}$$

which in the long time limit becomes

$$P_{\text{back}}(x) = \frac{x_0}{x} P_{\text{for}}(x), \tag{38}$$

where we used the property that in a steady state, the population growth rate and the individual growth rate are equal when there is no individual growth rate variability.

In some setups, experiments do not start with a unique ancestor cell but with $N_0 > 1$ initial cells, with possibly heterogeneous sizes. We describe this heterogeneity by the average initial size $\langle x_0 \rangle$ and the standard deviation $\sigma_{x_0}$. In this case, accessible sizes are still discrete but depend on both the number of divisions and the initial cell that started the lineage, and are expressed as $x_0^i \exp[\nu t]/2^K$, where $K$ takes integer values from 0 to $\infty$ and where $x_0^i \in \mathcal{X}_0$, with $\mathcal{X}_0$ the set of initial sizes. Consequently, a final size $x$ can possibly be reached by different couples $(K_i, x_0^i)$.

In order to go further, we now introduce explicitly the initial sizes $x_0^i$ in Eq. (34) as

$$\begin{aligned} h(x) &= \frac{1}{t} \ln \left[ \sum_K \sum_i 2^K R_{\text{for}}(K, x_0^i|x) \right] \\ &= \frac{1}{t} \ln \left[ \sum_K \sum_i 2^K R_{\text{for}}(K|x, x_0^i) R_{\text{for}}(x_0^i|x) \right]. \end{aligned} \tag{39}$$

When conditioning on the initial size $x_0^i$, there is only one possible number of divisions $K$ to reach size $x$, so that $R_{\text{for}}(K|x, x_0^i)$ obeys an equation similar to Eq. (35).

Let us examine two limit cases: (i) small variability in the initial sizes and (ii) large variability in the initial sizes.

Case (i) is characterized by a small number $N_0$ of initial cells and a small coefficient of variation $\sigma_{x_0}/\langle x_0 \rangle$. In this case, it is realistic to say that a final size $x$ can only be reached by one couple $(K^*, x^*)$, because the sets of accessible sizes generated by each initial cell do not overlap. Therefore, $R_{\text{for}}(x_0^i|x) = \delta(x_0^i - x^*)$ and so for any final size $x$, only one initial size $x^*$ survives in the sum, so that Eq. (39) reads $h(x) = \nu + \ln \left( x^*/(x^* \exp[\nu t]/2^K) \right)/t = \tilde{h}(K)$. Thus cells that come from lineages with the same number of divisions $K$ have the same fitness landscape value $h(x)$ for the size, regardless of the size $x^*$ of the initial cell of their lineages. Thus, available values for $h(x)$ are quantified by $K$ and form plateaus, where points representing cells coming from different ancestors but with the same number of divisions accumulate, as shown in Fig. 4a.

Case (ii) is characterized by a large number $N_0$ of initial cells and a large coefficient of variation $\sigma_{x_0}/\langle x_0 \rangle$. Unlike in case (i), the sets of accessible sizes generated by each initial cell have many overlaps, so that a final size $x$ can be reached by many different couples $(K_i, x_0^i)$. We make the hypothesis that a final size $x$ can be reached by any initial cell with uniform probability, so that $R_{\text{for}}(x_0^i|x) = 1/N_0$. Therefore, Eq. (39) becomes

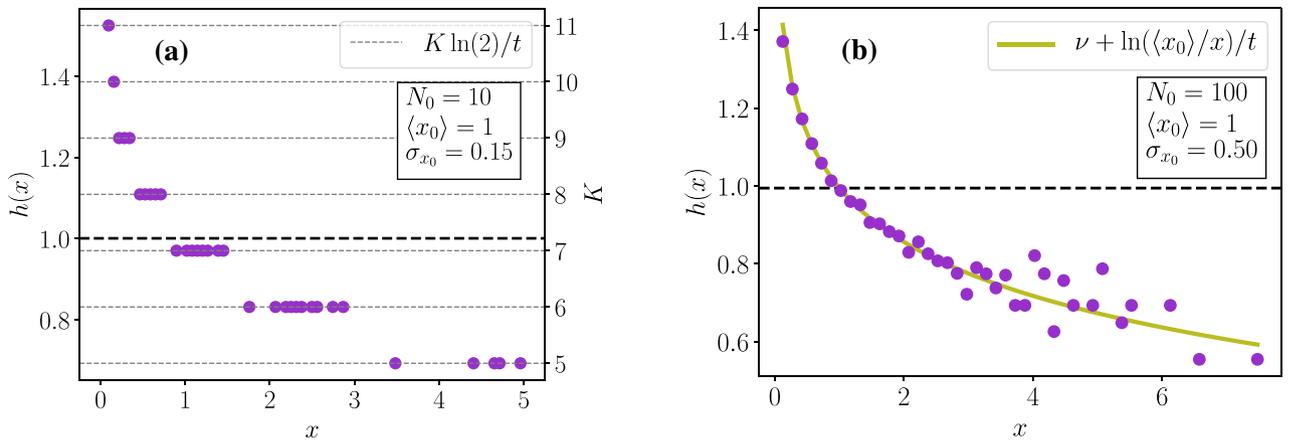

**Figure 4.** Size fitness landscapes for size models with $B(x) = \nu x^\alpha$, constant $\nu = 1, \alpha = 1, t = 5$, symmetrical division and $N_0$ initial cells following a Gaussian distribution of sizes $\mathcal{N}(\langle x_0 \rangle, \sigma_{x_0})$. The black horizontal dashed lines represent the population growth rates $\Lambda$. In (**a**), there is small variability in initial sizes: $N_0 = 10$, $\sigma_{x_0}/\langle x_0 \rangle = 0.15$. The grey dotted lines represent the plateaus available for $h(x)$, depending on the number of divisions $K$. The right y-axis displays the values of $K$ corresponding to these plateaus. In (**b**), there is large variability in initial sizes: $N_0 = 100$, $\sigma_{x_0}/\langle x_0 \rangle = 0.5$. The green curve is the theoretical prediction, on which the simulated purple dots align.

$$h(x) = \frac{1}{t} \ln \left[ \frac{1}{N_0} \sum_i \frac{x_0^i e^{\nu t}}{x} \right]$$
$$= \nu + \frac{1}{t} \ln \frac{\langle x_0 \rangle}{x}. \qquad (40)$$

This behavior was tested numerically and the result plotted on Fig. 4b confirms that the plateaus observed in case (i) are replaced by a smooth curve depending on the mean initial size.

We observe the same effect, namely the loss of the plateaus, when introducing fluctuations in individual growth rates.

**Age models.** *Constant individual growth rate.* We consider the case where the individual growth rate is constant and equal to $\nu$. In steady-state, the forward age distribution reads (see[17] where $p_{\text{for}}(a)$ (resp. $p_{\text{back}}(a)$) were denoted $p(a)$ (resp. $P(a)$)):

$$p_{\text{for}}(a) = p_{\text{for}}(0) \exp\left[-\int_0^a B(a')da'\right]. \qquad (41)$$

To find the integration constant $p_{\text{for}}(0)$, we use the normalization of probability $p_{\text{for}}$:

$$Z = p_{\text{for}}(0)^{-1} = \int_0^\infty da \exp\left[-\int_0^a B(a')da'\right]. \qquad (42)$$

Similarly, the steady-state backward distribution of ages reads

$$p_{\text{back}}(a) = p_{\text{back}}(0) \exp\left[-\Lambda a - \int_0^a B(a')da'\right]. \qquad (43)$$

In this case, the integration constant $p_{\text{back}}(0)$ can be expressed both using the normalization of $p_{\text{back}}(a)$, as done for the forward case, or using $p_{\text{back}}(0) = 2\Lambda$, as shown in[17].

Therefore, the ratio of the age distributions using the backward and forward statistics reads

$$\frac{p_{\text{back}}(a)}{p_{\text{for}}(a)} = 2Z\Lambda e^{-\Lambda a}, \qquad (44)$$

where $Z$ is defined in Eq. (42) and only depends on the division rate $B(a)$. This relation has a similar form as the relation derived by Hashimoto et al.[12] for the distributions of generation times, except for the extra age-independent factor $Z\Lambda$. Finally, the fitness landscape reads

$$h(a) = \frac{1}{t}\left[\Lambda(t - a) + \ln(2Z\Lambda)\right]. \qquad (45)$$

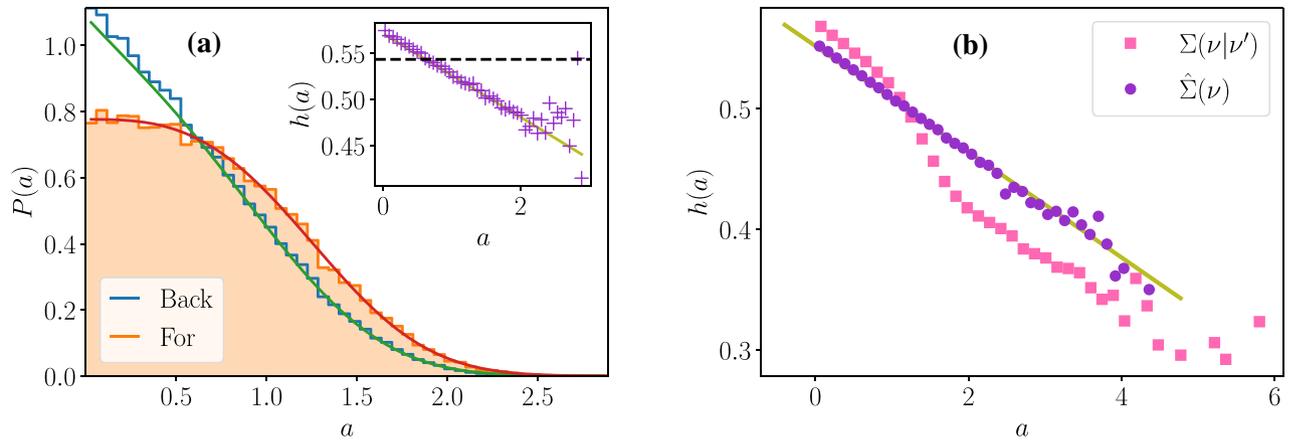

**Figure 5.** Age distributions and age fitness landscape for constant individual growth rate $\nu$ (**a**) and age fitness landscapes for fluctuating $\nu$ (**b**). In (**a**), the forward (resp. backward) age distribution is shown with orange filled histogram (resp. blue empty histogram). The red and green curves are the corresponding theoretical predictions. The inset plot shows the age fitness landscape (purple crosses) and the theoretical linear law (green). The horizontal black dashed-line represents the population growth rate $\Lambda$. The simulation was conducted with $B(a) = \nu a^\alpha$, $\nu = 1$, $\alpha = 2$, $t = 12$ and $N_0 = 1$. In (**b**), the age fitness landscapes are shown for models without (purple dots) and with (pink squares) mother-daughter correlations in individual growth rates. The green line of slope $-\Lambda$ fits well the purple dots. For both models, we chose a Gaussian distribution for $\Sigma$, of standard deviation $\sigma_\nu$, and centred either on the mother growth rate $\nu'$ for models with correlations or on a constant mean growth rate $\nu_m$ for uncorrelated models. The simulation was conducted with $\nu_m = 1$, $\sigma_\nu = 0.4$, $\alpha = 5$, $t = 12$ and $N_0 = 1$.

For the same reason as for $h(x)$ in size models, $h(a)$ in age models is a decreasing function of $a$ because lineages that divided a lot are over-represented in the population and are therefore more likely to contain young cells at time $t$.

The initial condition does not play any role in this derivation, therefore, unlike size models, the results obtained are unchanged for any number $N_0$ of initial cells with heterogeneous initial ages.

The above calculation is general because we did not put any constraint on $B(a)$. Let us now go into more details by choosing a power law for the division rate: $B(a) = \nu a^\alpha$. In this case, the integral of Eq. (42) is solvable and gives

$$Z = \frac{1}{\alpha+1}\left(\frac{\alpha+1}{\nu}\right)^{\frac{1}{\alpha+1}} \Gamma\left(\frac{1}{\alpha+1}\right), \tag{46}$$

in terms of Gamma function $\Gamma(x)$. Results are plotted on Fig. 5a, which shows that theoretical predictions for the backward and forward age distributions are in good agreement with the numerical histograms. The inset plot shows the age fitness landscape, which follows the linear behavior predicted by Eq. (45).

Let us examine the particular case of uncontrolled models, for which the division rate is constant: $B = \nu$. This corresponds to the case $\alpha = 0$ in the power law analysis conducted above. Replacing $\alpha$ by 0 in Eq. (46) leads to $Z = 1/\nu$; moreover in steady state $\Lambda = \nu$, so that

$$p_{\text{back}}(a) = 2 p_{\text{for}}(a) e^{-\Lambda a}. \tag{47}$$

Moreover, the distributions themselves are greatly simplified and read

$$p_{\text{for}}(a) = \nu e^{-\nu a}, \tag{48}$$

$$p_{\text{back}}(a) = 2\nu e^{-2\nu a}, \tag{49}$$

which shows that in this special case the age distributions are themselves identical with the generation time distributions.

*Fluctuating individual growth rates.* Another interesting extension of this calculation concerns the case of fluctuating individual growth rate $\nu$, for which the division rate then becomes a function of $a$ and $\nu$: $B(a, \nu)$. Then, steady state age distributions are[17]:

$$p_{\text{for}}(a) = \int d\nu\, p_{\text{for}}(0, \nu) \exp\left[-\int_0^a B(a', \nu) da'\right], \tag{50}$$

$$p_{\text{back}}(a) = e^{-\Lambda a} \int d\nu \, p_{\text{back}}(0, \nu) \exp\left[-\int_0^a B(a', \nu) da'\right], \tag{51}$$

where $p_{\text{for}}(0, \nu)$ and $p_{\text{back}}(0, \nu)$ are given by the boundary conditions:

$$p_{\text{for}}(0, \nu) = \int da d\nu' B(a, \nu') \Sigma(\nu|\nu') p_{\text{for}}(a, \nu'), \tag{52}$$

$$p_{\text{back}}(0, \nu) = 2 \int da d\nu' B(a, \nu') \Sigma(\nu|\nu') p_{\text{back}}(a, \nu'). \tag{53}$$

In the absence of mother-daughter correlations for the individual growth rate, then $\Sigma(\nu|\nu') = \hat{\Sigma}(\nu)$, which implies that $p_{\text{for}}(0, \nu)$ and $p_{\text{back}}(0, \nu)$ have the same dependency in $\nu$:

$$p_{\text{for}}(0, \nu) = \hat{\Sigma}(\nu) \int da d\nu' B(a, \nu') p_{\text{for}}(a, \nu'), \tag{54}$$

$$= \hat{\Sigma}(\nu) \hat{Z}^{-1} \tag{55}$$

$$p_{\text{back}}(0, \nu) = 2\hat{\Sigma} \int da d\nu' B(a, \nu') p_{\text{back}}(a, \nu') \tag{56}$$

$$= 2\hat{\Sigma}(\nu) \Lambda. \tag{57}$$

Finally, the fluctuation relation for the age reads

$$\frac{p_{\text{back}}(a)}{p_{\text{for}}(a)} = 2\hat{Z}\Lambda e^{-\Lambda a}, \tag{58}$$

which is the equivalent of Eq. (44) for fluctuating growth rates without mother-daughter correlations. Therefore, the age fitness landscape features the same linear dependency in age with a slope $-\Lambda$ as in the case of constant individual growth rate.

In the general case with mother-daughter correlations, this statement is not necessarily true though, because $p_{\text{for}}(0, \nu)$ and $p_{\text{back}}(0, \nu)$ do not have in general the same dependency in $\nu$.

Consequently, looking at the slope of the age fitness landscape informs on the presence of mother-daughter correlations as illustrated numerically in Fig. 5b, where the age fitness landscape for models without mother-daughter correlations aligns with the theoretical prediction of slope $-\Lambda$; while the same function for models with mother-daughter correlations presents a non-linear age dependency.

## Discussion

We have studied the relation between two different samplings of lineages in a branched tree: one sampling called backward or retrospective presents a statistical bias with respect to the forward or chronological sampling, an observation which is important to relate experiments carried out at the population level with the ones carried out at the single lineage level. This statistical bias can be rationalized by a set of fluctuation relations, which relate the probability distributions in the two ensembles and which are similar to fluctuation relations known in Stochastic Thermodynamics. This analogy leads to efficient methods to infer the population growth rate from an analysis of lineages, as we demonstrated by the analysis of the mother machine data of Tanouchi et al.[20]. Important inequalities for the mean number of divisions or the mean generation times follow from these fluctuation relations, which have been verified experimentally[12] for various strains of E Coli. It would be interesting to generalize these studies to other cell types, and in the particular context of this paper, it would be useful to perform experimental studies in bulk or semi open configurations, to test the predictions which involve a comparison between forward and backward samplings.

By measuring the difference between these two samplings for a specific trait, one obtains the fitness landscape, introduced by Nozoe et al.[13]. While these authors have applied that concept to variables which are not reset or redistributed at division in their work, in the present paper, we used the concept of fitness landscape for variables like the size and the age, which precisely undergo a reset at division in size and age models. We derived expressions for these fitness landscapes, which agree with the statistical bias which we expect when measuring size or age distributions in cell populations. In addition, we also find that the precise form of the age fitness function appears to inform whether or not mother-daughter correlations are present in age models.

In the future, it would be valuable to extend our approach to include other important phenotypic state variables besides size or age, such as variables controlling replication dynamics[3,27]. We hope that our work contributes to clarifying the connection between single lineage and population statistics and to understanding the fundamental constraints which cell growth and division must obey.

### Acknowledgements
The authors acknowledge R. García-García for a previous collaboration, which made possible the present work. We would also like to thank L. Robert, P. Gaspard and J. Unterberger for stimulating discussions.

### Author contributions
D.L. and A.G. designed the study and wrote the manuscript. A.G. conducted the calculations and numerical simulations, and analyzed the data.

### Competing interests
The authors declare no competing interests.

### Additional information
**Correspondence** and requests for materials should be addressed to A.G.

**Reprints and permissions information** is available at www.nature.com/reprints.

**Publisher's note** Springer Nature remains neutral with regard to jurisdictional claims in published maps and institutional affiliations.